\documentclass[twocolumn,superscriptaddress,fleqn,showpacs,showkeys,byrevtex,letterpaper]{revtex4}
\usepackage{amsfonts,amsmath,amssymb}
\usepackage{epsfig}
\usepackage{graphics}
\usepackage{dcolumn}
\usepackage{bm}

\begin{document}
\title{The Ferromagnetic Potts model under an external
magnetic field: an exact renormalization group approach}
\author{S\'ergio  Coutinho}
\email{sergio@lftc.ufpe.br} \affiliation{Departamento de
F\'{\i}sica, Universidade Federal de Pernambuco, CEP 50670-901,
Recife, Pernambuco, Brazil.}
\author{Welles A.\ M.\ Morgado}
\affiliation{Departamento de F\'{\i}sica, Pontif\'{\i}cia
Universidade Cat\'olica do Rio de Janeiro, CP 38071, CEP 22452- 970,
Rio de Janeiro, Brazil.}
\author{Evaldo M.\ F.\ Curado}
\affiliation{Centro Brasileiro de Pesquisas F\'{\i}sicas, Rua Xavier
Sigaud 150, CEP 22290-180, Rio de Janeiro, Brazil.}
\author{Lad\'ario da Silva}
\affiliation{Escola Naval, Marinha do Brasil, Av. Almte. Silvio de
Noronha s/n, CEP 20021-010, Rio de Janeiro, Brazil.}

\begin{abstract}
The q-state ferromagnetic Potts model under a non-zero magnetic
field coupled with the $0^{th}$ Potts state was investigated by an
exact real-space renormalization group approach. The model was
defined on a family of diamond hierarchical lattices of several
fractal dimensions $d_F$. On these lattices, the renormalization
group transformations became exact for such a model when a
correlation coupling that singles out the $0^{th}$ Potts state was
included in the Hamiltonian. The rich criticality presented by the
model with $q=3$ and $d_F=2$ was fully analyzed. Apart from the
Potts criticality for the zero field, an Ising-like phase transition
was found whenever the system was submitted to a strong reverse
magnetic field. Unusual characteristics such as cusps and
dimensional reduction were observed on the critical surface.
\end{abstract}

\maketitle

\section{Introduction}

The $q$-state Potts model is one of the most studied
models in statistical mechanics due to its wide
theoretical interest and practical applications
\cite{potts52,wu82,dorogovtsev04,igloy02,day05,baxter05a,
baxter82,bretz77,lubensky78,robinson82,healy85,parks93}.
The pure ferromagnetic version of this model in a zero
field is known to exhibit a phase transition from the
high-temperature disordered paramagnetic phase to the
low-temperature ferromagnetic for $q > 1$ and Bravais
lattices with dimensions $d \geq 2$ \cite{wu82,baxter82}.

From a theoretical viewpoint, the q-state Potts model in the absence
of an external magnetic field and defined on many kinds of Bravais
and fractal lattices, has been extensively studied by
mean-field-like and numerical approaches for ferromagnetic,
anti-ferromagnetic and disordered couplings. However, few exact
results have been obtained for such
models~\cite{baxter82,ladario96,timonin04,baxter05b}. Exact results
are very important as a guide for gaining an insight into many
complicated points of the phase diagram.

The q-state Potts model under an external magnetic field in its
turn, due to its complexity, has been studied by only a few authors
\cite{kassan03,karsch00,kim04,glumac94} and several aspects of its
physical behavior have remained poorly established.

This paper has studied the role of the magnetic field on the
physical properties of the ferromagnetic q-state Potts model defined
on a family of fractal lattices, called diamond hierarchical
lattices (DHL). The zero-field version of this model was previously
studied by the authors \cite{ladario96}. One important issue for the
study of spin models on such lattices arises from the scale
invariance property which is presented and that leads to exact and
tractable solutions. Generally speaking, in qualitative terms, they
compare favorably to similar ones obtained by other approximate
methods. Indeed, the study of spin models on such lattices can be
viewed as the counterpart of the real-space renormalization group
approximation for the corresponding models defined on Bravais
lattices \cite{migdal75,kadanoff76}. Therefore, they provide an
alternative framework for studying the critical behavior of more
realistic systems whenever approaches considering translational
invariant lattices lead to untractable procedures.

To benefit from the scale invariance property of the DHL, an exact
real-space renormalization group (RG) scheme was applied in order to
study the q-state Potts model on such lattices. Closed
renormalization transformations in the coupling parameter space were
achieved when a $0$-state correlation coupling interaction was
introduced into the Hamiltonian, as well as the constants of the
magnetic field and ferromagnetic coupling. Under such $0$-state
correlation interaction pairs of spins in the $0^{th}$ Potts state,
aligned with the magnetic field, are singled out. The particular
model with $q=3$ Potts states defined on a DHL with a scaling factor
of two and fractal dimension of $d_F=2$ was exhaustively studied.
The full flow diagram of the closed RG equations in the three
dimensional parameter space was obtained exhibiting the stable fixed
points associated with the zero-field ferromagnetic and paramagnetic
Potts phases, as well as those associated with an Ising-like phase
induced by a strong reverse magnetic field. The critical surface
delineating the basin of attraction of the Ising-like fixed point
was sketched, revealing unusual characteristics such as cusps and
dimensional reduction.

This paper is organized as follows: the Hamiltonian model and the
general renormalization equations for the model with $q$ Potts
states defined on a general DHL with an arbitrary fractal dimension
are presented and discussed in section \ref{sectionII}, and its
deduction is displayed in the Appendix. The particular model for
$q=3$ and $d_F=2$ is studied in section \ref{sectionIII}, the
corresponding renormalization equations set in subsection
\ref{subsectionIIIA}, while subsection \ref{subsectionIIIB} is
devoted to presenting the critical points in the coupling constant
parameter space, and to analyzing the full renormalization flow and
unusual characteristics appearing on the critical surface. Finally,
the discussion is presented and the conclusions are summarized in
section \ref{sectionV}.

\section{The Hamiltonian model and
methodology}\label{sectionII}

\begin{figure}
\centering
\includegraphics[height=6.5cm]{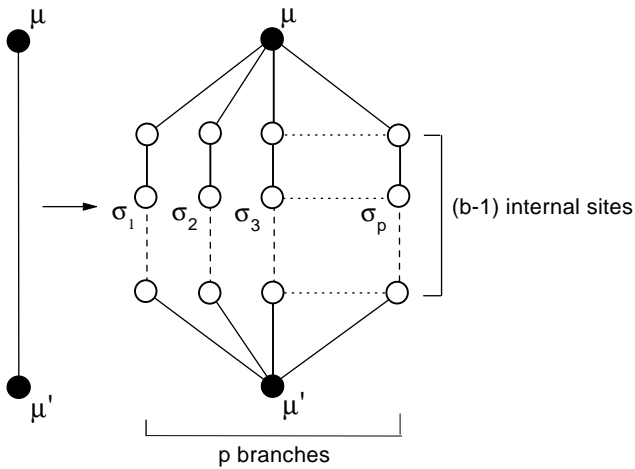}
\caption{Inflation process for a general diamond
hierarchical
lattice with scaling factor $b$ and $p$ branches.}
\label{figure01}
\end{figure}

DHL lattices were constructed commencing from a basic unit (called
first generation or hierarchy) and replaced all of its single bonds
with the basic unit itself to build the second generation. Such a
replacing procedure is sketched in Figure \ref{figure01} for the
general DHL, with a scaling factor $b$ and $p$ connecting branches,
hereafter referred to as $(b,p)$-DHL. This inflation procedure is
recursively applied leading to a two-root lattice with the fractal
dimension, the number of sites and the number of bonds given by
$d_F=1+\log p/\log b$, $N_{S}^{n}=2+(b-1)p[(bp)^{n}-1]/(bp-1)$ and
$N_{B}^{n}=(bp)^{n}$, respectively, with $n$ as the number of
generations or hierarchies.

The general $q$-state Potts Hamiltonian
$\mathcal{H}^{(n)}$ on a $n$-generation $(b,p)$-DHL can
be written as:
\begin{multline}\label{eq01}
-\beta \mathcal{H}^{(n)} = \sum_{<i,j>}
qK^{(n)}_{ij}\delta_{\sigma_i\,\sigma_j} +\\
+\sum_{i}qH^{(n)}_{i}\delta_{\sigma_i\,0}+
\sum_{<i,j>}qL^{(n)}_{ij}\delta_{\sigma_i\,0}\delta_{\sigma_j\,0},
\end{multline}
where $\beta=1/kT$, $T$ is the absolute temperature and
$\delta_{\sigma,\sigma'}$ is the Kronecker delta function. The
energy of the ferromagnetic interaction coupling the
nearest-neighbor spins $\sigma$'s ($\sigma=0,\,1,\,2,\dots q$) is
described by the first term of the equation (\ref{eq01}), while the
energy of the interaction of the external magnetic field aligned to
the $0^{th}$ state is given by the second term. Finally, the above
mentioned correlation coupling interaction, which bolsters the
energy of the pairs of spins in the $0^{th}$-state is accounted for
by the third term.

The Hamiltonian function for the pure (homogeneous)
ferromagnetic Potts model obtained from equation
(\ref{eq01}) when all coupling constant interactions are
considered independent of the site position, that is $
K^{(n)}_{ij} \equiv K^{(n)}$, $L^{(n)}_{ij}
\equiv L^{(n)}$, and $H^{(n)}_{ij} \equiv H^{(n)}$, is
simplified as
\begin{multline}\label{eq02}
-\beta \mathcal{H}^{(n)} = qK^{(n)}\sum_{<i,j>}
\delta_{\sigma_i\,\sigma_j} +\\ +
qH^{(n)}\sum_{i}\delta_{\sigma_i\,0}+
qL^{(n)}\sum_{<i,j>}\delta_{\sigma_i\,0}\delta_{\sigma_j\ ,0}
\end{multline}

The pure $q$-state Potts model, defined on the $(b,p)$-DHL with $n$
generations, can be exactly transformed by renormalization in an
equivalent model defined on the $(b,p)$-DHL with $(n-1)$ generation
as long as the $p(b-1)$ internal spin variables within each basic
unit are decimated.

The new Hamiltonian coupling constants are obtained by a formal set
of renormalization equations given by
\begin{gather*}
K^{(n-1)} = F_1 (K^{(n)},L^{(n)},H^{(n)}), \\
H^{(n-1)} = F_3 (K^{(n)},L^{(n)},H^{(n)}),\\
L^{(n-1)} = F_2 (K^{(n)},L^{(n)},H^{(n)}),\\
\end{gather*}
\noindent where the functions $F_1$, $F_2$ and $F_3$, in
general, are determined by the decimation process.

From this point, appropriate compact variables
(\textit{transmissivities}) introduced by Tsallis and Levy
\cite{tsallis81} were considered.
\begin{gather}
t_{n} =\frac{e^{qK^{(n)}}-1}{e^{qK^{(n)}}+(q-1)} ,
\label{eq03} \\
u_{n} = \frac{e^{qH^{(n)}}-1}{e^{qH^{(n)}}+(q-1)},\label{eq04} \\
v_{n} =\frac{e^{qL^{(n)}}-1}{e^{qL^{(n)}}+(q-1)}
\label{eq05}
\end{gather}
The equivalent decimation relations for the above variables ($t$,
$u$, $v$), rather than ($K$, $L$, $H$) were obtained for the
particular $(2,p)$-DHL family with the scaling factor $b=2$. The
choice of such hierarchical lattice family (scaling factor $b=2$) is
appropriated for studying the ferromagnetic model. To investigate
the antiferromagnetic model, for instance, hierarchical lattices
with odd scaling factor (odd $b$) should be required to preserve the
ground-state configuration under renormalization. The deduction of
the $b=2$ decimation relations is presented in the Appendix and
leads formally to closed renormalization-group coupled equations
given by
\begin{gather}
t_{n-1}=\frac{T(t_n,u_n)-1}{T(t_n,u_n)+q-1},
\label{eq06}\\
u_{n-1}=\frac{U(t_n,u_n,v_n)-1}{U(t_n,u_n,v_n)+q-
1}, \label{eq07} \\
v_{n-1}=\frac{V(t_n,u_n,v_n)-1}{V(t_n,u_n,v_n)+q-1} \label{eq08}
\end{gather}
where
\begin{gather}
T(t,u)=\left[1+\frac{qt^2(1-u)}{(1-t)(1+t-2tu)}\right]^p,
\label{eq09}\\
U(t,u,v)=\left[\frac{1+(q-1)u}{1-
u}\right]\left[\frac{N_2}{D}\right]^p, \label{eq10}\\
V(t,u,v)=\left[\frac{N_3\,D^2}{N_1\,N_2^2}\right]^p \label{eq11}
\end{gather}
with
\begin{gather}
D=\frac{q( 1 + t - 2tu )}{(1-t) ( 1-u ) }, \label{eq12}\\
N_1=\frac{qt[ 2 + (q-2 ) t ] }{( 1-t )^2} +\frac{q}{1-
u},\label{eq13}\\
N_2=q-2 + \frac{1 + ( q-1 ) t}{1 - t} - \nonumber \\- \frac{[ 1 + (
q-1 ) t ][ 1 + ( q-1 ) u ] [ 1 + ( q-1 ) v ] }{( 1 - t )
( 1-u ) ( 1 - v ) },\label{eq14}\\
N_3=q-1 + \nonumber \\+ \frac{[ 1 + (q -1) t ]^2 [ 1 + ( q-1 ) u ] [
1 + (q-1) v ]^2}{( 1 - t )^2( 1 - u ) ( 1 - v )^2}.\label{eq15}
\end{gather}
Previous results for the zero field pure $q$-state Potts model were
re-obtained \cite{ladario96} when the limits of the  zero field
(u=0) and the zero correlation coupling (v=0) limits are imposed on
the equations (\ref{eq06})-(\ref{eq08}), that is
\begin{equation*}
      t_{n-1}=\frac{2t_n^2+(q-2)t_n^4}{1+(q-1)t_n^4}.
\end{equation*}

\section{The $q=3$-state Potts model on a (2,2)-DHL with fields}\label{sectionIII}

The particular model with $q=3$ and $b=p=2$, was investigated in the
present Section. Such model, which corresponds to the simplest one
defined on a hierarchical lattice with integer fractal dimension
$d_F > 1$, is suitable for comparision with the corresponding model
defined on the square lattice in the framework of the real-space
renormalization group approximation~\cite{migdal75,kadanoff76}. Even
though it presents a rich criticality with unusual characteristics
on the critical surface. The study of other cases should follow
straightforward the present calculation although with much heavy
algebraic computational efforts.

\subsection{The renormalization transformations}\label{subsectionIIIA}

The renormalization equations for the $3$-state Potts model defined
on the $(2,2)$-DHL, which has the fractal dimension $d_F=2$, written
in terms of the transmissivity variables, can be directly obtained
from (\ref{eq06})- (\ref{eq15}) by setting $q=3$ and $p=2$. After
some algebraic rearrangements the following coupled renormalization
equations are obtained:
\begin{gather}
      t_{n-1}=\frac{T(t,u)-1}{T(t,u)+2}, \label{eq16}\\
      u_{n-1}=\frac{U(t,u,v)-1}{U(t,u,v)+2},
\label{eq17}\\
      v_{n-1}=\frac{V(t,u,v)-1}{V(t,u,v)+2} \label{eq18}
\end{gather}
where
\begin{gather}
T(t,u)=\frac{[1+(2-u)t^2-2tu]^2}{(1-t)^2[1+(1-2u)t]^2}, \label{eq19}
\end{gather}
\begin{multline}\label{eq20}
U(t,u,v)=\frac{( 1 + 2u)}{1-u}\times \\ \times \frac{[ 1 + 2uv + t(
1 + u + v + 3uv )]^2 }{[ 1 +t( 1 - 2u)]^2( 1 - v)^2},
\end{multline}
\begin{multline}
V(t,u,v)=\frac{W(t,u,v)}{9[(1-t)^2 + t(2+t)(1-u) ]^2}\times\\
\times \frac{[ 1 + t( 1 - 2u ) ]^4}{[ 1 + 2uv +t( 1 + u + v +
3uv)]^4}, \label{eq21}
\end{multline}
\begin{multline}
W(t,u,v)=[ 2{( 1 - t ) }^2( 1 - u ) ( 1 - v )^2 +  \\ + ( 1 + 2t )
^2 ( 1 + 2u ) ( 1 + 2v )^2] ^2. \label{eq22}
\end{multline}

\subsection{Renormalization Flow Diagram and Critical
Points:
}\label{subsectionIIIB}

The 3D-renormalization parameter space in the variables $(t,u,v)$
has the advantage of being confined to the region ($-0.5\leq t\leq
1;\ -0.5\leq u\leq 1;\ -0.5\leq v\leq1)$ the bounds corresponding to
the $\pm \infty$ values of the respective former variables ($K$,\,
$H$,\, $L$). However the region $-0.5\leq t<0$ corresponding to
$K<0$, although allowed by the variable transformation (\ref{eq03})
must be excluded by the following conceptual reason: the $(2,p)$-DHL
family (scaling factor two) is not suitable for studying the
antiferromagnetic systems ($K<0$) since the RG decimation process
does not preserve the antiferromagnetic ground state symmetry under
renormalization. DHL with an odd scaling factor must be considered
in order to study antiferromagnetic models within real space RG
approaches.

\begin{figure}[h]
\includegraphics[width=80mm]{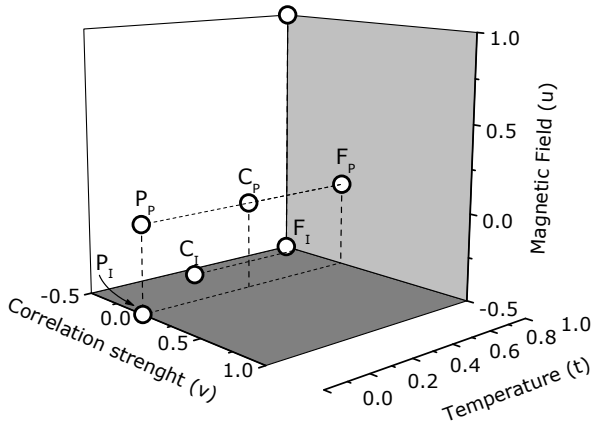}\\
\caption{3D-plot displaying the fixed points described in
Table~\ref{table}. $\textsf{P}_\textsf{P}$ and
$\textsf{F}_\textsf{P}$ denote the fixed points associated with
paramagnetic and ferromagnetic Potts phases, while
$\textsf{C}_\textsf{P}$ label the unstable fixed point.
$\textsf{P}_\textsf{I}$, $\textsf{F}_\textsf{I}$ and
$\textsf{C}_\textsf{I}$ indicate the corresponding fixed points
associated with the Ising-like transition, respectively.
}\label{figure02}
\end{figure}

The fixed points of the corresponding recursion equations
(\ref{eq16})-(\ref{eq18}) are shown in Table~\ref{table} and
sketched in Figure~\ref{figure02}.
\begin{table}[!]
\centering
\begin{tabular}{|c|c|c|l|}
\hline
    \multicolumn{4}{|c|}{Zero-field Potts model:
$H=L=0$ or line $u=v=0$.}\\ \hline
    t & u & v & description \\ \hline
    0 & 0 & 0 & Potts paramagnetic phase \\
    $\frac{1}{2}$ & 0 & 0 & Potts transition \\
    1 & 0 & 0 & Potts ferromagnetic phase \\ \hline
    \multicolumn{4}{|c|}{Negative infinite field $H=-
\infty$ or plane $u=-0.5$ }\\ \hline
    t & u & v & description \\ \hline
    0 & $-\frac{1}{2}$& 0 & Ising paramagnetic phase  \\
   $0.442687\dots$\footnote{Exact value$=-\left(
\frac{4}{19} \right)  +
  \frac{{\left( 17766 - 3078\,{\sqrt{33}} \right)
}^{\frac{1}{3}}}{57} +
  \frac{{\left[ 2\,\left( 329 + 57\,{\sqrt{33}} \right)
\right] }^
     {\frac{1}{3}}}{19}$} & $-\frac{1}{2}$ & $-
0.397513\dots$ & Ising like transition \\
    1 & $-\frac{1}{2}$&$-\frac{1}{2}$  & Ising
ferromagnetic phase \\ \hline
    \multicolumn{4}{|c|}{Zero temperature invariant
plane: $T=0$ or plane $t=1.$}\\ \hline
    t & u & v &  description\\ \hline
    1 & 1 & $-\frac{1}{2}$ &  \\
    1 & 0 & 0 & Potts ferromagnetic phase \\
    1 & $-\frac{1}{2}$&$-\frac{1}{2}$  & Ising
ferromagnetic phase \\ \hline
\end{tabular}
\caption{Fixed points of the renormalization flow diagram of the
$3$-state Potts Model under non-zero field. The $t,\,u,\,v$
variables are defined by equations (\ref{eq03})-(\ref{eq05}).}
\label{table}
\end{table}

\begin{figure}[h]
\includegraphics[width=80mm]{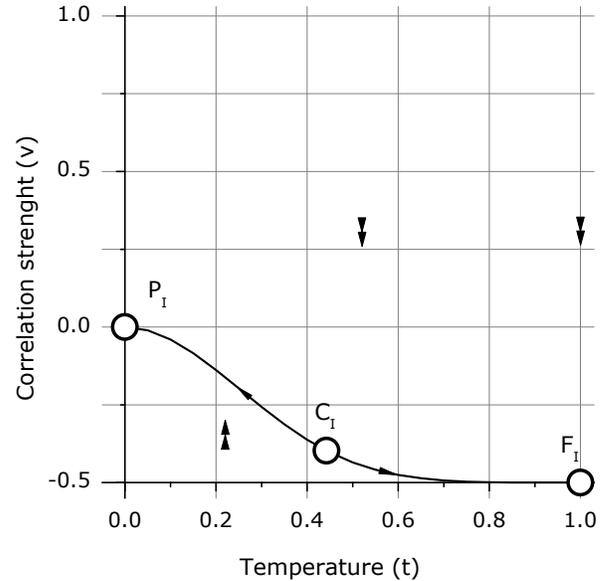}\\
\caption{Flow Diagram $(t \times v)$ corresponding to an Ising like
model in the invariant plane $u=-0.5$. The open circles mark the
loci of the fixed points. $\textsf{P}_\textsf{I}$,
$\textsf{F}_\textsf{I}$ label the stable-fixed points of the
paramagnetic (P) and ferromagnetic (F) phases while
$\textsf{C}_\textsf{I}$ labels the unstable-fixed point (critical
point). The arrows indicate the flow direction, the double arrows
signifying a direct jump to the flow line (solid
line).}\label{figure03}
\end{figure}

\begin{figure}[tb]
  \includegraphics[width=80mm]{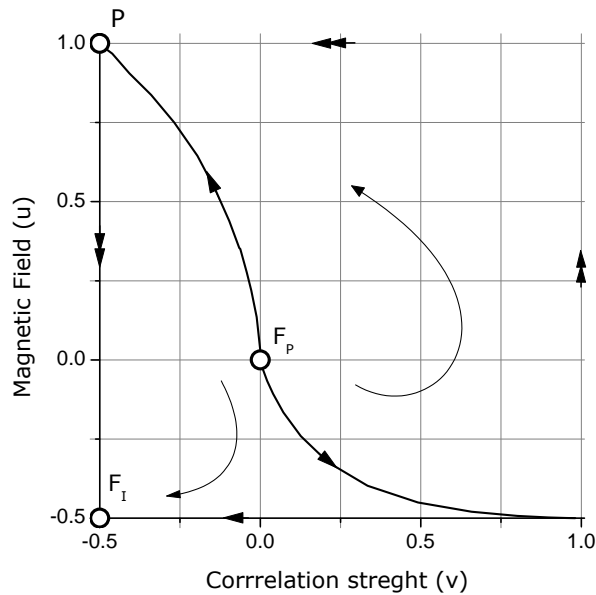}\\
  \caption{Flow Diagram $(v \times u)$ corresponding to
the zero temperature invariant plane $t=1$. Open symbols mark the
fixed points, $\textsf{F}_\textsf{P}$ labels the stable fixed point
of the Potts ferromagnetic phase, $\textsf{F}_\textsf{I}$ labels the
stable fixed point of the Ising ferromagnetic phase, while
\textsf{P} indicates the stable fixed-point associated paramagnetic
phase (infinite $H$ field). The arrows indicate the flow direction,
the double arrows signifying a direct jump to the end point. The
solid plot corresponds to the intersection of the critical surface
with the $t=1$ plane. Note the singularity in the derivative of the
plot at the point $\textsf{F}_\textsf{P}$ (0,0).}\label{figure04}
\end{figure}

Three manifolds in the 3D cubic parameter space $(t,u,v)$
can be distinguished:
\begin{description}
\item[(a)] The zero-field and zero-correlation coupling
Potts model fixed points located on the line $u=v=0$,
which contains the stable fixed points (0,0,0) and
(1,0,0) associated with the paramagnetic and the
ferromagnetic phases respectively,  as well as the
unstable fixed point $(0.5,0,0)$ associated with the
corresponding phase transition, the latter obtained by
the exact solution of $t^{*}=2 t^{* 2}/ (1+2t^{* 4})$.
\item[(b)]The $u = -1/2$ ($H \rightarrow -\infty$) plane, which
corresponds to the Ising model without magnetic field. The reduction
of the number of states from $q=3$ to $q=2$ occurred because the
$0$-state became unreachable. For the general $q$-state Potts
ferromagnetic model, whenever $u = -1/2$ $(H \rightarrow -\infty)$
the $0$-state becomes unreachable, leading to a $(q-1)$-state Potts
model without magnetic field on such manifold. The renormalization
flow in the $u=-0.5$ plane is depicted in Figure \ref{figure03}. The
Ising-like paramagnetic ($0,-0.5,0$) and the ferromagnetic
(1,-0.5,-0.5) stable fixed points as well as the unstable phase
transition points ($0.442687\dots,\,-0.5,\,-0.397513\dots$) are
found within this plane. Exact values may be obtained as indicated
in Table \ref{table}. The renormalization flow is confined in the
plane $u=-0.5$ and given by
\begin{gather}
 t'=\frac{3\,t^2\,{\left( 2 + t \right) }^2}
  {4 + 8\,t - 4\,t^3 + 19\,t^4}\label{eq23}\\
  v'=\frac{2A-B}{A+B}\label{eq24}
\end{gather}
with
\begin{gather*}
A=16[(1+2t)(1-t)]^4\\
B=[2(1-t)^2+3(2+t)]^2(2+t)^4.
\end{gather*}
\item[(c)] The zero temperature invariant plane $(t=1)$,
which contains both the Potts and the Ising fixed points,
$(1,0,0)$ and $(1,-0.5,-0.5)$ respectively, is displayed
in the Figure~\ref{figure04}.
\end{description}

\begin{figure}
  \includegraphics[width=85mm]{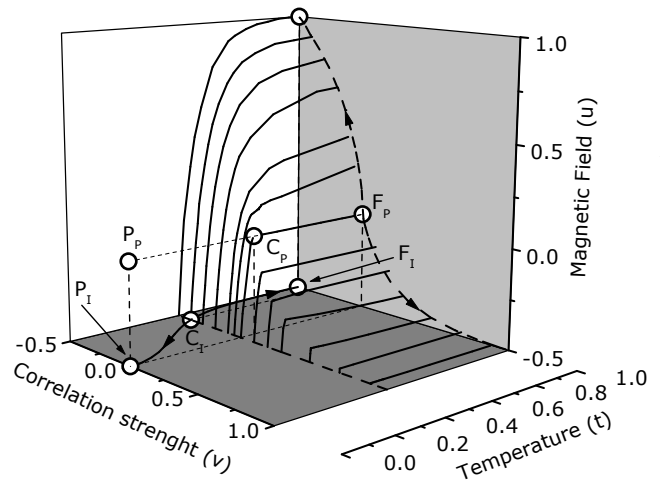}\\
  \caption{Profiles of intersections of the critical
surface with the planes $v=-0.5$, $-0.395$, $-0.3$, $-0.2$, $-0.1$,
$-0.05$, $0.0$, $0.1$, $0.2$, $0.3$, $0.5$, $0.7$ and  $0.9$ (solid
lines), from top to bottom. Open circles mark the loci of the fixed
points, $\textsf{F}_\textsf{P}$, $\textsf{P}_\textsf{P}$,
$\textsf{C}_\textsf{P}$ label ferromagnetic, paramagnetic and Potts
transition fixed points, while \textsf{F}$_\textsf{I}$,
\textsf{P}$_\textsf{I}$ and $\textsf{C}_\textsf{I}$ indicate the
corresponding Ising like fixed points at the plane
$u=-0.5$.}\label{figure05}
\end{figure}
\section{Discussion and conclusions}\label{sectionV}
The critical properties of the $q$-state ferromagnetic Potts model
defined on a fractal hierarchical lattice in the presence of a
magnetic field, were investigated through a real-space
renormalization group (RG) technique. The inclusion of a new
interaction term in the Hamiltonian that boosted the coupling of
pairs of spins in the $0^{th}$ state (the state chosen to be singled
out by the external field) led to a closed -- and exact -- coupled
RG transformation. The full renormalization flow diagram in an
appropriate compact parameter space is analyzed for the particular
model with $q=3$ Potts states, defined on the diamond hierarchical
lattice with a fractal dimension $d_F=2$. The inclusion of such a
correlation coupling breaks the symmetry for the pair correlation
with respect to the Potts state singled out by the external field.
Whenever $q\geq 3$ pairs of spins with states parallel to the
external field have different energies with respect to all pairs of
parallel spins in other states. Note that such a correlation
coupling has no relevance for the Ising model ($q=2$) leading only
to rescale the energy of the ferromagnetic pair interaction.

\begin{figure}
  \includegraphics[width=85mm]{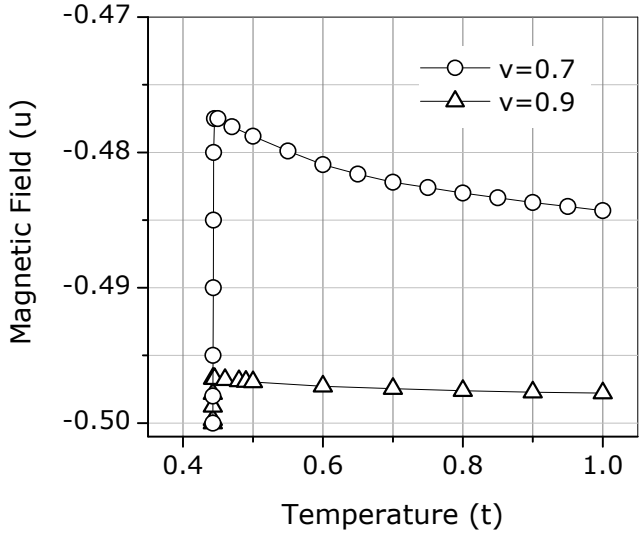}\\
  \caption{Details of intersections of the critical
surface with the planes $v=0.7$ and  $v=0.9$, from top to bottom.
The vertical scale was expanded to emphasize the cusps for lower
values of $t$.}\label{figure06}
\end{figure}

New and interesting features are found within the global
phase diagram:
\begin{description}
\itemsep=-1mm
      \item[(a)] A dimensional reduction of the order
parameter at a strong (infinite) reverse magnetic field,
which depopulates the $0^{th}$ Potts state: within the
plane ($u=-1/2$) the system is reduced to a $(q-1)$-Potts
like model. For the studied case ($q=3$) the reduced
model corresponds to an effective Ising ferromagnet.
Within the 3D-parameter space flow diagram the stable
fixed point associated with the Ising condensed phase is
located at zero temperature ($t=1$) and infinite reverse
magnetic field ($u=-1/2$).
      \item[(b)] The basin of attraction of such a point
is separate from that of the paramagnetic phase by a
complex boundary surface, which contains the locus of the
zero field $q$-state Potts phase transition. Both the
unstable and the stable fixed points associated with the
Potts phase belongs to the boundary surface along the
line ($u=v=0$, $t \geq 1/2$)
      \item[(c)] Along such a line, the boundary surface
has a sharp cusp. Other lines with a sharp cusp also appears in the
$v\geq 0$ portion of the surface.
\end{description}
Figure \ref{figure05} displays several plots of the intersection of
the critical surface with planes ($t,u$) for chosen values of $v$,
showing all the above-mentioned features. Figure~\ref{figure06}
exhibits in appropriated scale the details of sharp cusps appearing
in the plots of the intersection of the critical surface with the
planes $v=0.7$ and $v=0.9$. Whether such features are present in the
$q$-state Potts model defined in other kinds of lattices deserves
further investigations.

The eigenvalues and eigenvectors associated with the Potts and Ising
unstable fixed points were calculated. Table~\ref{table2} displays
such result. The $\alpha$ and $\nu$ critical exponents at the fixed
points associated with the Potts and Ising transitions, previously
reported in the literature\cite{ladario96} were recovered. The
general nature of the transitions studied in this work (for $q=3$)
is found to be of a second order. Other authors~\cite{timonin04}
have claimed that, at higher values of $q$, the inherent
inhomogeneity of the model might induce a change to first-order
transition, which provides another question for further
investigation.
\begin{table}[!]
\centering
\begin{tabular}{|c c c c l|} \hline
 \multicolumn{5}{|c|}{Potts transition point: (0.5,\ 0,\ 0)}\\ \hline
$\lambda_t=$ & $2.26938$&$\rightarrow$ & $\vec{v}_t=$ &  $(-0.18388,\ 0.813562,\ 0.551639)$ \\
$\lambda_u=$ & $ 1.77777$&$\rightarrow$ & $\vec{v}_u=$ & $(1,\ 0,\ 0)$ \\
$\lambda_v=$ & $1.17506$&$\rightarrow$ & $\vec{v}_v=$ &
$(-0.159264,\ -0.863917,\ 0.477792)$
\\ \hline
\multicolumn{5}{|c|}{Ising transition point: (0.442687,\ -0.5,\ -0.397513)} \\
\hline $\lambda_t=$ & $1.678573$&$\rightarrow$ & $\vec{v}_t=$ & $(0.912186,\  0,\  -0.409776)$ \\
$\lambda_u=$ & $0.419643$&$\rightarrow$ & $\vec{v}_u=$ &  $(0.0499605,\  0.99851,\ -0.0219619)$ \\
$\lambda_v=$ & $0$&$\rightarrow$ & $\vec{v}_v=$ & $(0,\ 0,\ 1)$ \\
\hline
\end{tabular}
\caption{Eigenvalues $\lambda$'s and the corresponding eigenvectors
$\vec{v}$'s associated with the transtion Potts and Ising fixed
points.} \label{table2}
\end{table}

In closing, to the current knowledge of the authors, the
study reported throughout this paper is one of a few
exact renormalization group approaches to the Potts model
under an external magnetic field in a growing body of
literature on the subject.\\

\begin{acknowledgments}
Work partially supported by Brazilian granting agencies
CNPq, CAPES,
FAPERJ and FACEPE.
\end{acknowledgments}

\appendix*

\section{Exact renormalization equations for the $q$-state Potts
model on $(2,p)$-DHL}\label{appendix}

The Hamiltonian for a single basic unit with one internal site $b=2$
can be written as:
\begin{multline}\label{eqA01}
-\beta \mathcal{H} =
qK\sum_{i=1}^p(\delta_{\sigma_i\,\mu_1}+\delta_{\sigma_i\ ,\mu_2})
+qL\sum_{i=1}^p(\delta_{\sigma_i\,0}\delta_{\mu_1\,0}+\\
+\delta_{\sigma_i\,0}\delta_{\mu_2\,0})+
qH\sum_{i=1}^p[\delta_{\sigma_i\,0} +
\delta_{\mu_1\,0}+\delta_{\mu_2\,0}],
\end{multline}
which can be rewritten for latter convenience as
\begin{equation*}
 -\beta \mathcal{H} =\sum_{i=1}^p \mathcal{H}_i
\end{equation*}
with
\begin{multline}\label{eqA02}
\mathcal{H}_i=qK(\delta_{\sigma_i\,\mu_1}+\delta_{\sigma_ i\,\mu_2})
+qL(\delta_{\sigma_i\,0}\delta_{\mu_1\,0}
+\delta_{\sigma_i\,0}\delta_{\mu_2\,0})+\\ + qH
\delta_{\sigma_i\,0}+\frac{q}{p}H\delta_{\mu_1\,0}+\frac{
q}{p}H\delta_{\mu_2\,0}.
\end{multline}
The corresponding restricted partition function for fixed
configuration of the root sites:
\begin{multline}\label{eqA03}
z(\mu_1,\mu_2)=Tr_{\{\sigma_i\}}e^{-\beta \mathcal{H}} =
Tr_{\{\sigma_i\}}\prod_{i=1}^p e^{\mathcal{H}_i}=\\=
\left\{Tr_{\sigma_i}e^{\mathcal{H}_i}
\right\}^p=\left\{\sum_{\sigma=0}^{q-
1}\exp\left[qK(\delta_{\sigma\,\mu_1}+\delta_{\sigma\,\mu _2})
+\right.\right.\\+
\left.\left.qL(\delta_{\sigma\,0}\delta_{\mu_1\,0} +
 \delta_{\sigma\,0}\delta_{\mu_2,0})+
\frac{q}{p}H\,(p\,\delta_{\sigma\,0} +\right.\right.\\+ \left.\left.
\delta_{\mu_1\,0}+\delta_{\mu_2\,0})\right]\right\}^p.
\end{multline}
The renormalized bond Hamiltonian is defined by
\begin{equation}\label{eqA04}
-\beta \mathcal{H}' = qK' \delta_{\mu_1\,\mu_2}
+qL'\delta_{\mu_1\,0}\delta_{\mu_2\,0}+ qH'(
\delta_{\mu_1\,0}+\delta_{\mu_2\,0}),
\end{equation}
while the renormalized restricted partition function results as:
\begin{multline}\label{eqA05}
z'(\mu_1,\mu_2)=e^{-\beta \mathcal{H}'} =\exp\left[qK'
\delta_{\mu_1\,\mu_2}
+\right.\\\left.+qL'\delta_{\mu_1\,0}\delta_{\mu_2\,0}+ qH'(
\delta_{\mu_1\,0}+\delta_{\mu_2\,0})\right].
\end{multline}
The partition functions $Z'$ and $Z$ are compared to obtain the
renormalization equations for the coupling constants $K',\ L',\ H'$
as a function of $K,\ L,\ H$, that is
\begin{gather}
      Z\equiv A\,Z', \nonumber \\
      \sum_{\{\mu_1,\mu_2\}}z(\mu_1,\mu_2)=A\,
      \sum_{\{\mu_1,\mu_2\}}z'(\mu_1,\mu_2),\label{eqA06}
\end{gather}
where $A$ is a constant to be determined as a function of $K,\ L,\
H$, which can be used to fix the origin of the energy scale. The
expansions of the configurations of the root sites in the sums of
equation~(\ref{eqA06}) are performed and four kinds of possible
configurations may be distinguished, namely
\begin{enumerate}
\item $\mu_1=\mu_2=0$, one configuration.
\item $\mu_1=0$ and $\mu_2\neq 0$ and \textit{vice-
versa}, two configurations.
\item $\mu_1=\mu_2\neq 0$, leading to $(q-1)$
configurations.
\item $\mu_1\neq \mu_2$,  $\mu_1\mu_2\neq 0$,
leading to $(q-1)(q-2)$ configurations.
\end{enumerate}
The contributions for the partition functions are equal by symmetry
for each case. The comparison term by term led to the following
relations:
\begin{gather}
A\,e^{qK'+qL'+2qH'} = e^{2qH}(e^{2qK+2qL+qH}+q-
1)^p,\label{eqA07}\\
A\,e^{qH'}=e^{qH}(e^{qK}+e^{2qH+qL}+q-2)^p,\label{eqA08}\\
A\, e^{qK'}=(e^{qH}+e^{2qK}+q-2)^p,\label{eqA09}\\
A=(e^{qH}+2e^{qK}+q-3)^p,\label{eqA10}
\end{gather}
which can be solved  for the coupling constants $K',\ L',\ H'$
leading to
\begin{gather}
      e^{qK'}=\left(\frac{e^{2qK}+e^{qH}+q-
2}{2e^{qK}+e^{qH}+q-3}\right)^p,\label{eqA11}\\
      e^{qH'}=e^{qH}\left(\frac{e^{qK}+e^{qK+qL+qH}+q-
2}{2e^{qK}+e^{qH}+q-3}\right)^p,\label{eqA12}
\end{gather}
\begin{multline}
 e^{qL'}=\left[\frac{(e^{2qK+2qL+qH}+q-
1)}{(e^{2qK}+e^{qH}+q-2)}\times \right.\\ \left. \times
\frac{(2e^{qK}+e^{qH}+q-3)^2}{(e^{qK}+e^{qK+qL+qH}+q-2)^2}\right]^p.\label{eqA13}
\end{multline}
New appropriated compact variables $t,\,u,\,v$, defined in the
interval $(-1/(q-1),1)$ instead of the $K,\,H,\,L$ defined in
$(-\infty,\infty)$, are considered respectively by
\begin{gather}
      t=\frac{e^{qK}-1}{e^{qK}+q-1}, \label{eqA14}\\
      u=\frac{e^{qH}-1}{e^{qH}+q-1}, \label{eqA15}\\
      v=\frac{e^{qL}-1}{e^{qL}+q-1}. \label{eqA16}
\end{gather}
The inverse relations give
\begin{gather}
      e^{qK}=\frac{1+(q-1)t}{1-t}, \label{eqA17}\\
      e^{qH}=\frac{1+(q-1)u}{1-u}, \label{eqA18}\\
      e^{qL}=\frac{1+(q-1)v}{1-v}. \label{eqA19}
\end{gather}
Equations (\ref{eqA17})-(\ref{eqA19}) are substituted in
(\ref{eqA11})-(\ref{eqA13}), which by it turns are substituted  in
the equations (\ref{eqA14})-(\ref{eqA16}) after changing
($t\rightarrow t'$). The result is
\begin{gather}
      t'=\frac{T(t,u)-1}{T(t,u)+q-1}, \label{eqA20}\\
      u'=\frac{U(t,u,v)-1}{U(t,u,v)+q-1}, \label{eqA21}\\
      v'=\frac{V(t,u,v)-1}{V(t,u,v)+q-1}, \label{eqA22}
\end{gather}
where
\begin{gather}
T(t,u)=\left[1+\frac{qt^2(1-u)}{(1-t)(1+t-2tu)}\right]^p,
\label{eqA23}\\
U(t,u,v)=\left[\frac{1+(q-1)u}{1-
u}\right]\,\left[\frac{N_2}{D}\right]^p,\label{eqA24}\\
V(t,u,v)=\left[\frac{N_3\,D^2}{N_1\,N_2^2}\right]^p \label{eqA25}
\end{gather}
with
\begin{gather}
D=\frac{q( 1 + t - 2tu )}{(1-t) ( 1-u ) },
\label{eqA26}\\
N_1=\frac{qt[ 2 + (q-2 ) t ] }{( 1-t )^2} + \frac{q}{1-u},
\label{eqA27}\\
N_2=q-2 + \frac{1 + ( q-1 ) t}{1 - t} - \nonumber \\
-\frac{[ 1 + (q-1 ) t ][ 1 + ( q-1 ) u ]
[ 1 + ( q-1 ) v ] }{( 1 - t ) ( 1-u ) ( 1 - v )},\label{eqA28}\\
N_3=q-1 + \nonumber \\+\frac{[ 1 + (q -1) t ]^2 [ 1 + ( q-1 ) u ] [1
+ ( q-1 ) v ]^2}{( 1 - t )^2( 1 - u ) ( 1 - v )^2}.\label{eqA29}
\end{gather}
Equations (\ref{eqA20})-(\ref{eqA29}) give the complete set of the
renormalization equations for the $q$-state Potts Model defined on a
general diamond hierarchical lattice with scaling factor $b=2$ and
fractal dimension $d_F=1+\log p/\log 2$, whose Hamiltonian is given
by equation (\ref{eq02}).

\end{document}